\documentclass[prd,floatfix,twocolumn,amsmath,amssymb,floatfix,nofootinbib]{revtex4}
\usepackage{graphicx,color,dcolumn,booktabs,bm}
\usepackage{longtable,lscape}
\usepackage{txfonts}
\usepackage{overpic}
\usepackage{epsfig}
\usepackage{amssymb}
\usepackage{rotating}
\usepackage{epstopdf}
\usepackage{indentfirst}
\usepackage{feynmf}   
\usepackage{slashed}  
\usepackage{cases}
\usepackage{color}
\usepackage{multirow}
\usepackage{graphicx,color,dcolumn,booktabs,bm}

\graphicspath{{Figures/}} %

\graphicspath{{Figures/}} %
\usepackage[colorlinks, citecolor=blue,anchorcolor=red,menucolor=red, linkcolor=red,filecolor=red,runcolor=red,frenchlinks=red]{hyperref}
\usepackage{cases}

\begin{document}

\title{Puzzle of the $\Lambda_c$ spectrum}

\author{Qi-Fang L\"u$^{1}$}\email{lvqifang@ihep.ac.cn}
\author{Yubing Dong$^{1,2}$}\email{dongyb@ihep.ac.cn}
\author{Xiang Liu$^{3,4}$}\email{xiangliu@lzu.edu.cn}
\author{Takayuki Matsuki$^{5,6}$}\email{matsuki@tokyo-kasei.ac.jp}
\affiliation{
$^1$Institute of High Energy Physics, CAS, Beijing 100049, China\\
$^2$Theoretical Physics Center for Science Facilities (TPCSF), CAS, China\\
$^3$School of Physical Science and Technology,
Lanzhou University,
Lanzhou 730000, China\\
$^4$Research Center for Hadron and CSR Physics, Lanzhou University
$\&$ Institute of Modern Physics of CAS,
Lanzhou 730000, China\\
$^5$Tokyo Kasei University, 1-18-1 Kaga, Itabashi, Tokyo 173-8602, Japan\\
$^6$Theoretical Research Division, Nishina Center, RIKEN, Wako, Saitama 351-0198, Japan}

\begin{abstract}

There is a puzzle in the $\Lambda_c^+$ family, i.e., one member with $J^P=3/2^+$ is missing in a $L=2$ multiplet which the heavy quark effective theory predicts, and $J^P$'s of $\Lambda_c(2765)^+$ and $\Lambda_c(2940)^+$ are unknown.
Using a light diquark picture to calculate baryon masses, we study possible assignments of two $\Lambda_c$'s with unknown $J^P$ and the missing $\Lambda_c^+$ with $3/2^+$ for $L=2$, and we find the most probable possibility that the peak corresponding to $\Lambda_c(2880)^+$ actually includes a missing member with spin $3/2^+$ for $L=2$ and that quantum numbers of $\Lambda_c(2765)^+$ and $\Lambda_c(2940)^+$ are $2S(1/2^+)$ and $2P(1/2^-)$, respectively.

\end{abstract}

\maketitle

{\it Introduction:} In the former paper \cite{Matsuki:2016hzk}, we have pointed out that careful observation of the experimental spectra of heavy-light mesons tells us that heavy-light mesons with the same angular momentum $L$ are almost degenerate and that mass differences within a heavy quark spin doublet and between doublets with the same $L$ are very small compared with a mass gap between different multiplets with different $L$, which is nearly equal to the value of the $\Lambda_{QCD}\sim 300$ MeV. In Conclusions and Discussion of Ref. \cite{Matsuki:2016hzk}, we have also suggested that $\Lambda_c^+$ baryons may have properties similar to heavy-light mesons.
There have been a couple of papers which pursue the similar idea for light and vector mesons by Afonin and his collaborators \cite{Afonin:2007aa,Afonin:2013hla}, on which we do not discuss in this paper.

According to PDG \cite{Agashe:2014kda}, there are six $\Lambda_c^+$ baryons observed by experiments, which are $\Lambda_c(2286)^+$, $\Lambda_c(2595)^+$, $\Lambda_c(2625)^+$, $\Lambda_c(2765)^+$ (or $\Sigma_c(2765)^+$), $\Lambda_c(2880)^+$, and $\Lambda_c(2940)^+$. Among these, the following heavy quark spin multiplets are identified; $\Lambda_c(2286)$ with $J^P=1/2^+$ for $L=0$, $\Lambda_c(2595)^+$ and $\Lambda_c(2625)^+$ with $1/2^+$ and $3/2^+$ for $L=1$, respectively, and $\Lambda_c(2880)^+$ with $5/2^+$ for $L=2$. There is one member missing in the spin multiplet for $L=2$ which has spin $3/2^+$. Other than this missing particle, mass differences in the same multiplet, i.e., with the same $L$, are very small and gaps between the average masses of spin multiplets are all $\sim300$ MeV, which obeys the rule proposed in Ref. \cite{Matsuki:2016hzk}. Although these successful assignments, there still remains a puzzle in the $\Lambda_c^+$ baryons, where the missing member for $L=2$ is and what quantum numbers $J^P$ are for $\Lambda_c(2765)^+$ and $\Lambda_c(2940)^+$. If we regard $\Lambda_c(2940)^+$ as a missing member of $L=2$, then this state must have spin-parity $3/2^+$, which contradicts a common understanding that a state with smaller spin $3/2$ appears lower in mass than a state with larger spin $5/2$. { In addition, the strong and radiative decays of $\Lambda_c(2940)$ in a $D^* N$ molecular scenario have been analyzed in Refs.~\cite{Dong:2009tg,Dong:2010xv} which are against spin 3/2}.

There is a pioneering work \cite{Capstick:1986bm} which calculates baryon masses directly extending the method of Godfrey and Isgur \cite{Godfrey:1985xj}.
However, since this method is complicated, in this article regarding a baryon as a heavy quark-light diquark system ($Q\{qq\}$)  like in Ref. \cite{Ebert:2011kk} and calculating its mass, we show that the above observation on $\Lambda_c$ holds, predict $J^P$ and spin assignments of $\Lambda_c(2765)^+$ and $\Lambda_c(2940)^+$, and propose a solution to a puzzle where the missing member for $L=2$ is.

We adopt the method provided in Ref. \cite{Ebert:2011kk} to calculate baryon masses whose prescription is given by: 1) First we calculate diquark masses assuming the relativised quark model proposed by Godfrey and Isgur (GI) \cite{Godfrey:1985xj}. 2) Next, having a diquark mass calculated and regarding two light quarks as ${3}^*\in{3}\times {3}$, we again apply the GI model to a heavy quark-light diquark system to obtain a baryon mass.

This way of calculation places very tight conditions on parameters so that it is very difficult to reproduce physical masses as one can imagine. Hence, we also refer to the values of baryon masses where diquark masses are parameters to fit with experimental data \cite{Chen:2016iyi}, and also refer to the values given in Refs. \cite{Capstick:1986bm,Ebert:2011kk,Roberts:2007ni}.\\

{\it Relativized Quark Model and Diquark Masses:}
To calculate baryon masses, we adopt interactions propopsed by the relativized GI model whose Hamiltonian between quark and antiquark can be expressed as
\begin{equation}
\tilde{H} = H_0+\tilde{V}(\vec{p},\vec{r}), \label{ham}
\end{equation}
where
\begin{eqnarray}
H_0 &=& (p^2+m_1^2)^{1/2}+(p^2+m_2^2)^{1/2},\\
\tilde{V}(\boldsymbol{p},\boldsymbol{r}) &=& \tilde{H}^{conf}_{12}+\tilde{H}^{cont}_{12}+\tilde{H}^{ten}_{12}+\tilde{H}^{so}_{12}.
\end{eqnarray}
Here $\tilde{H}^{conf}_{12}$ includes the spin-independent linear confinement and Coulomb-like interaction. $\tilde{H}^{cont}_{12}$, $\tilde{H}^{ten}_{12}$, and $\tilde{H}^{so}_{12}$ are the color contact, color tensor interactions, and spin-orbit term, respectively. Subindices 1 and 2 denote quark (${\bf 3_c}$) and antiquark (${\bf 3^*_c}$), respectively. The symbol ``$\sim{}$'' on top of the operator $\tilde{H}$ means that we operate the relativized procedure on $H$, by which relativistic effects are taken into account. The explicit forms of those interactions and the details of the relativization procedure can be found in Refs.~\cite{Godfrey:1985xj,Capstick:1986bm} for mesons and baryons, respectively.

To solve Eq. (\ref{ham}), we need values of parameters which are given in Table \ref{table1} provied by Refs. \cite{Godfrey:1985xj,Capstick:1986bm}. Here $C_{qq}$ is taken to be the same as $C_{q\bar q}$ in Ref. \cite{Godfrey:1985xj} up to the factor, an inner product of color matrices $\left<\vec F_1\cdot\vec F_2\right>$, with $V_{string}(\vec{r})=C_{qq} ({\rm ~or~}C_{q\bar q})+br$. $\tilde V$ includes Gell-Man matrices whose expectation vlaues, $\left<\vec F_1\cdot \vec F_2 \right>$, are {\-4/3} for $q\bar q$ and -2/3 for a diquark $qq$. In Table \ref{table1}, GI means parameters taken from Ref. \cite{Godfrey:1985xj} and CI from Ref. \cite{Capstick:1986bm} which we adopt in this work to calculate diquark masses as well as heavy quark-diquark, i.e., charmed baryon masses. In this work, $C_{q\bar q}$ is used for $C_{Qdi}$ where $Q$ is a heavy quark and ``di'' expresses a diquark. The word ``same'' in the CI column in Table \ref{table1} means the same value as a GI parameter.

\begin{table}[htbp]
\begin{center}
\caption{Values of parameters used in this work. GI means parameters taken from Ref. \cite{Godfrey:1985xj} and CI from Ref. \cite{Capstick:1986bm}. \label{table1}}
\footnotesize
\begin{tabular}{lcc}
\hline\hline

                             & GI   & CI (This work)       \\ \hline
$\frac{1}{2}(m_u+m_d)$~(MeV) & 220                &  same  \\
$m_s$~(MeV)                  & 419                &  same  \\
$m_c$~(MeV)                  & 1628               &  same  \\
$\alpha_s$                   & 0.60               &  same  \\
$\Lambda_{\rm QCD}$~(MeV)    & 200                &  same  \\
$C_{q\bar q}$~(MeV)          & ${-\frac{4}{3}(-253)}$ & same  \\
$C_{qq}$~(MeV)               & N/A                & ${-\frac{2}{3}(-253)}$\\
$\sigma_0$~(GeV)             & 1.80 & same                 \\
$s$                          & 1,55 & same                 \\
b~(GeV$^2$)                  & 0.18 & 0.15                 \\
$\frac{1}{2}+\epsilon_{\rm cont}$    & $\frac{1}{2}-0.168$ & same                \\
$\frac{1}{2}+\epsilon_{\rm tens}$    & $\frac{1}{2}+0.025$ & $\frac{1}{2}-0.168$ \\
$\frac{1}{2}+\epsilon_{{\rm so}(v)}$ & $\frac{1}{2}+0.055$ & $\frac{1}{2}$       \\
$\frac{1}{2}+\epsilon_{{\rm so}(s)}$ & $\frac{1}{2}+0.055$ & $\frac{1}{2}+0.30$  \\
$\frac{1}{2}+\epsilon_{\rm Coul}$    & $\frac{1}{2}$       & same                \\

\hline\hline

\end{tabular}
\end{center}
\end{table}

The calculated diquark masses by using both parameter sets of GI and CI are listed in Table~\ref{table2}. Although Ref.~\cite{Capstick:1986bm} includes the three body interaction, we neglect it to simplify the calculation.

\begin{table}[!htbp]
\begin{center}
\caption{\label{table2} Masses of scalar and axial vector diquarks. $S$ and $A$ denote scalar and axial vector diquarks, respectively. The braces and brackets correspond to symmetric and antisymmetric quark contents in flavor, respectively. The units are in MeV.}
\small
\begin{tabular*}{8.5cm}{@{\extracolsep{\fill}}*{4}{p{2cm}<{\centering}}}
\hline\hline
 Quark content &  Diquark type & Mass (GI) & Mass (CI)\\\hline
 $[u,d]$       &  $S$            & 691               & 642\\
 $\{u,d\}$     &  $A$            & 840               & 779\\
 $[u,s]$       &  $S$            & 886               & 838\\
 $\{u,s\}$     &  $A$            & 992               & 934\\
 \hline\hline
\end{tabular*}
\end{center}
\end{table}

\begin{table*}[htbp]
\caption{\label{table3} Predicted masses for the $\Lambda_c^+$ state of ours and other approaches in Refs.~\cite{Ebert:2011kk,Chen:2016iyi,Capstick:1986bm,Roberts:2007ni} compared to experimental data~\cite{Agashe:2014kda} (in MeV). We adopt masses generated by the CI parameter set in this work.}
\renewcommand\arraystretch{1.2}
\begin{tabular*}{170mm}{@{\extracolsep{\fill}}lccccccc}
\toprule[1pt]\toprule[1pt]
\multirow{2}{*}{States}  & \multicolumn{7}{c}{$\Lambda_c^+$ baryons}  \\
\cline{2-8}
    & PDG~\cite{Agashe:2014kda}  &   Prediction (CI)  &   Prediction (GI)  & Ref.~\cite{Chen:2016iyi} & Ref.~\cite{Ebert:2011kk} & Ref.~\cite{Capstick:1986bm} & Ref.~\cite{Roberts:2007ni}  \\
\cline{1-8}
$\mid 1S, 1/2^+\rangle$ &  2286.86  &  2177     &  2267      &  2286   &  2286   &  2265   &  2268     \\
$\mid 2S, 1/2^+\rangle$ &  2766.6   &  2749     &  2891      &  2772   &  2769   &  2775   &  2791     \\
$\mid 3S, 1/2^+\rangle$ &           &  {3160}     &  {3345}      &  3116   &  3130   &  3170   &           \\
$\mid 1P, 1/2^-\rangle$ &  2592.3   &  2603     &  2736      &  2614   &  2598   &  2630   &  2625     \\
$\mid 1P, 3/2^-\rangle$ &  2628.1   &  2619     &  2739      &  2639   &  2627   &  2640   &  2636     \\
$\mid 1D, 3/2^+\rangle$ &           &  2930     &  3095      &  2843   &  2874   &  2910   &  2887     \\
$\mid 1D, 5/2^+\rangle$ &  2881.53  &  2919     &  3065      &  2851   &  2880   &  2910   &  2887    \\
$\mid 2P, 1/2^-\rangle$ &  2939.3   &  3030     &  3204      &  2980   &  2983   &  3030   &  2816    \\
$\mid 2P, 3/2^-\rangle$ &           &  3038     &  3203      &  3004   &  3005   &  3035   &  2830    \\
\bottomrule[1pt]\bottomrule[1pt]
\end{tabular*}
\end{table*}

{\it Baryon masses: }\label{sec4}
After obtaining diquark masses in Table \ref{table2}, we calculate baryon masses using Eq. (\ref{ham}) with both parameter sets of GI and CI. Although the mass values with the CI parameter set is better than those of GI, we list both results in Table \ref{table3} as ``Prediction (CI/GI)'' together with experimental data and other results given by Refs. \cite{Chen:2016iyi,Ebert:2011kk,Capstick:1986bm,Roberts:2007ni}. Among these results, Refs. \cite{Chen:2016iyi,Ebert:2011kk} use heavy quark-diquark picture and Ref. \cite{Chen:2016iyi} uses the semi-relativistic quark potential model which should be compared with our results. Since Ref. \cite{Chen:2016iyi} treats diquark masses as parameters, they obtain better fit with experimental data than ours. { Our self-consistent calculation of the $\Lambda_c$ baryon masses gives a rather higher values compared with other models. Even though our calculated value 2930 MeV is very close to $\Lambda_c(2940)$, it is natural to consider that our values 2930 and 2919 MeV should form a multiplet for $L=2$. In this case, we find that mass difference betweeen members of a $L=2$ multiplet is within $\sim$10MeV including other models. } Accordingly, we can see the similar tendency for all the models:
\begin{enumerate}
\item $\Lambda_c(2765)^+$ is identified as a $\mid 2S, 1/2^+\rangle$ state.
\item $\Lambda_c(2940)^+$ is identified as a $\mid 2P, 1/2^-\rangle$ state.
\item The observed peak of the $\Lambda_c(2880)^+$ assigned as $\mid~1D, 5/2^+\rangle$ actually includes a missing state $\mid 1D, 3/2^+\rangle$ because their predicted masses listed in Table \ref{table3} are so close to each other, within $\sim 10$ MeV, which could not be distinguished by experiments.
\end{enumerate}
Items 2 and 3 have been already pointed out by the paper of Ref. \cite{Chen:2016iyi}.
As for Item 3, experimental errors of the mass and width for $\Lambda_c(2880)^+$ are so small, $2881.50\pm 0.35$ and $5.8\pm 1.1$ given in Refs. \cite{Agashe:2014kda,Abe:2006rz,Aubert:2006sp}, respectively, that one cannot imagine that there is a missing particle hidden in the same peak at $\Lambda_c(2880)^+$. However, there are theoretical uncertainties as one can see from Table \ref{table3}. References \cite{Capstick:1986bm,Roberts:2007ni} give the same mass values for $\mid1D, 5/2^+\rangle$ and $\mid 1D, 3/2^+\rangle$, and Refs. \cite{Ebert:2011kk,Chen:2016iyi} including ours give masses within { $\sim 10$ MeV} so that considering theoretical errors, both states are most probably in one peak or it is very difficult to separate two states from the peak at 2880 MeV.\\


{\it Conclusions and discussion: }\label{sec4}
After many $XYZ$ particles have been discovered, people have paid attention to these particles to solve a question how they can be explained, either molecular state, or tetra-quark state or just a kinematical effect \cite{Liu:2013waa,Chen:2016qju,Lebed:2016hpi}. Now settling most of the $XYZ$ particles, people are now paying most of their energy to study heavy baryons, the cases that one or two or three quarks are heavy. To attack this problem is a bit hard compared with the $XYZ$ particles because baryons consist of three quarks and it is very hard to solve a three-body problem. A diquark picture makes the problem easier to calculate their mass spectrum as well as their decay behaviors with a help of the $^3P_0$ model as the paper \cite{Eichten:1993ub}. Also see the paper \cite{Chen:2016iyi} how to apply the method proposed by \cite{Eichten:1993ub} to baryons.

After starting from the simple observation that many of the heavy-light mesons have degenerate masses within a heavy quark spin multiplet, we could extend the idea to baryons. The first example was the $\Lambda_c$ baryons. However, there is the puzzle in this spectrum that there is a missing member in a $L=2$ heavy-quark spin multiplet.

In this article, we have studied the $\Lambda_c^+$ spectrum. To do so, we have calculated baryon masses taking a heavy quark-diquark picture for baryons and have compared with other theoretical as well as experimental values. To take a diquark picture for two light quarks, we have tried to be self-consistent, i.e., we have calculated diquark masses at first using the GI model with the CI parameter set, and using those diquark masses, we have computed baryon masses. After observing the obtained values together with former theoretical values, we have concluded that $\Lambda_c(2765)^+$ and $\Lambda_c(2940)^+$ are identified as $\mid 2S, 1/2^+\rangle$ and $\mid 2P, 1/2^-\rangle$ states, respectively and that a missing member of the $L=2$ heavy-quark spin multiplet is hidden in the peak around $\Lambda_c(2880)^+$.

Because our prediction on a missing member of the $L=2$ heavy quark multiplet depends on accuracy of experiments, future careful measurements on the $\Lambda_c^+$ spectrum by LHCb and the forthcoming BelleII is waited for to test our prediction.

\section*{Acknowledgements}

This project is partly supported by the National
Natural Science Foundation of China under Grant
No. 11475192, as well as supported by the
DFG and the NSFC through funds provided to the Sino-German
CRC 110 Symmetries and the Emergence of Structure in QCD.
This project is also partly supported by the National Natural Science Foundation of China under Grant Nos. 11305003, 11222547, 11175073, 11447604 and U1204115.
T. Matsuki wishes to thank the institute of High Energy Physics, Beijing, where part of this work was carried out.

\end{document}